\documentstyle[preprint,aps,epsfig]{revtex}
\tighten
\title{{\bf Charmless hadronic decays $B \to VV$ in the
Topcolor-assisted Technicolor model }}
\author{ Zhenjun Xiao \thanks{Email address: zjxiao@email.njnu.edu.cn}
and  Libo Guo\thanks{Email address: guolibo@email.njnu.edu.cn}\\
{\small  Department of Physics, Nanjing Normal University,
Nanjing, Jiangsu 210097 People's Republic of China.}}
\date{\today}
\newcommand{\beq}{\begin{eqnarray}}
\newcommand{\eeq}{\end{eqnarray}}

\newcommand{\aepa}{a_{\epsilon'}}
\newcommand{\aepb}{a_{\epsilon + \epsilon'}}

\newcommand{\mw}{ M_W }
\newcommand{\paa}{\pi_1^{\pm}}
\newcommand{\pbb}{\pi_8^{\pm}}
\newcommand{\pcc}{\tilde{\pi}^{\pm}}

\newcommand{\pcz}{\tilde{\pi}^0}
\newcommand{\pdd}{\tilde{H}^{\pm}}

\newcommand{\mpaa}{m_{\pi_1}}
\newcommand{\mpbb}{m_{\pi_8}}
\newcommand{\mpcc}{m_{\tilde{\pi}}}

\newcommand{\fpit}{F_{ \tilde{\pi}}}
\newcommand{\nceff}{N_c^{eff} }

\def\acp{{\cal A}_{CP}}

\newcommand{\non}{\nonumber\\ }

\newcommand{\obar}[1]{\shortstack{{\tiny (\rule[.4ex]{1em}{.1mm})} \\ [-.7ex] $#1$}}
\begin{document}
\maketitle
\begin{abstract}
Based on the effective Hamiltonian with the generalized factorization approach,
we calculate the branching ratios and CP asymmetries of $B \to VV$
decays in the Topcolor-assisted Technicolor (TC2) model.
Within the considered parameter space we find that: (a)
for the penguin-dominated $B \to K^{*+}\phi$ and $K^{*0}\phi$ decays,
the new physics enhancements to the branching ratios are around $40\%$;
(b) the measured branching ratios of $B \to K^{*+} \phi$ and $ K^{*0} \phi$ decays
prefer the range of $3 \lesssim \nceff \lesssim 5$;
(c) the SM and TC2 model predictions for the branching ratio ${\cal B}(B^+ \to \rho^+
\rho^0)$ are only about half of the Belle's measurement;
and (d) for most $B \to VV$ decays,  the new physics corrections on their CP asymmetries
are generally small or moderate in magnitude and
insensitive to the variation of $\mpcc$ and $\nceff$.
\end{abstract}


\vspace{.5cm}



\newpage
\section{ Introduction } \label{sec:1}

As is well-known, one of the main goals of B experiments is to find the
evidence or signals of new physics beyond the standard model (SM).
Precision measurements of B meson system can provide an insight into very
high energy scales via the indirect loop effects of new physics\cite{slac504,xiao20},
and offer a complementary probe to the searches for new
physics in future collider runs at the Tevatron, and the LHC,
and the future linear $e^+ e^-$ colliders.

Theoretically, the low energy effective Hamiltonian\cite{buchalla96a}
is our  basic tool
to study the B meson decays. The short-distance QCD corrected Lagrangian at next-to-leading order
(NLO) is available now, but we still do not know how to calculate
hadronic matrix elements from first principles.
The generalized factorization (GF) ansatz \cite{bsw87,ali9804,cheng98} is widely used in
literature \cite{ali9804,du97,ali9805,chen99}, and the resulted predictions are
basically consistent with the experimental measurements. But we also know that
non-factorizable contribution really exists and can not be neglected numerically
for many hadronic B decay channels. Two new approaches, called as the QCD
factorization \cite{bbns} and the perturbative QCD approach\cite{li02},
appeared recently  and played an important role in reducing
the uncertainties of the corresponding theoretical predictions\cite{bbns,li02,qcdf,pqcd}.

During the past three decades, many new physics models beyond the SM have been constructed.
The most popular one is certainly the minimal supersymmetric standard model (MSSM), another
alternative to break the electroweak symmetry is the Technicolor mechanism\cite{weinberg76}.
The Topcolor-assisted Technicolor (TC2) model\cite{hill95} is a viable model and
consistent with current experimental data\cite{buchalla96b,epjc10,epjc18}.
In paper\cite{epjc10} we calculated the new electroweak penguin contributions to the
rare K decays in the TC2 model.
In a recent paper\cite{epjc18}, we presented our systematic calculation of branching ratios
and CP-violating asymmetries for two-body charmless hadronic decays $B \to P P$,
$P V $ (the light pseudo-scalar (P) and/or vector(V) mesons ) in the framework of
TC2 model \cite{hill95}. It is natural to extend our study to the cases of $B \to VV$ decays.

$B \to VV$ decays have been studied frequently in the SM and new physics models,
for example, in Refs.\cite{ali9804,ali9805,atwood99,xiao01,he98}.
Three decay modes, $B \to K^{*+}\phi, K^{*0}\phi$ and $\rho^+ \rho^0$ decays,
have been measured recently by CLEO, BaBar and Belle Collaborations
\cite{prl86-3718,prl87-151801,belle-0113,belle-0255,prd65-051101}.
In this paper, we will concentrate on the new physics
effects on nineteen charmless $ B \to VV$ decays.

This paper is organized as follows. In Sec.~2, we give a brief introduction
for the TC2 model and examine the constraints on the parameter space of
the TC2 model. In Sec.~3, based on previous analytical calculations of
new penguin diagrams, we find the effective Wilson coefficients
$C_i^{eff}$ and effective numbers $a_i$ with the inclusion of new physics
contributions. In Sec.~4 and 5, we calculate and show the numerical results
of branching  ratios  and CP-violating asymmetries for all nineteen
$B \to VV$ decay modes, respectively. We focus on those measured decay modes.
The conclusions are included in the final section.

\section{ Basics of the TC2 model} \label{sec:2}

Apart from some differences in group structure and/or particle contents,
all TC2 models have the similar common features.
In this paper we chose the well-motivated
and most frequently studied TC2 model proposed by Hill \cite{hill95}
as the typical TC2 model to calculate the new physics contributions to the B decays in question.

In the TC2 model, there exist top-pions $\pcc$ and $\pcz$, charged
b-pions $\pdd$ and neutral b-pions $( \tilde{H}^0, \tilde{A}^0)$,
and the techni-pions $\paa$ and $\pbb$.
The couplings of top-pions to t- and b-quark can be written as
\cite{hill95}:
\beq
\frac{m_t^*}{ \fpit } \left[ i\bar{t} t \tilde{\pi}^0 +
  i \overline{t}_R b_L \tilde{\pi}^+
+ \frac{m_b^*}{m_t^*}  \overline{t}_L b_R \tilde{\pi}^+ + h.c.  \right]
\eeq
here, $m_t^* = (1-\epsilon) m_t$ and $m_b^* \approx 1 GeV$ denote the
masses of top and bottom quarks generated by topcolor interactions.
At low energy, potentially large FCNCs arise
when the quark fields are rotated from their weak eigenbasis to their mass
eigenbasis, realized by the matrices $U_{L,R}$ and $D_{L,R}$:
\beq
b_L \to  D_L^{bd} d_L  +   D_L^{bs} s_L + D_L^{bb} b_L, \\
b_R \to D_R^{bd} d_R + D_R^{bs} s_R + D_R^{bb} b_R,
\eeq
the FCNC interactions will be induced
\beq
\frac{m_t^*}{\fpit} \left[
i\tilde{\pi}^+ ( D_L^{bs}\bar{t}_R  s_L +  D_L^{bd}\bar{t}_R d_L) +
i\tilde{H}^+ ( D_R^{bs} \bar{t}_L s_R +   D_R^{bd}\bar{t}_L d_R)
+ h.c. \right ]
\eeq

For the mixing matrices in the TC2 model, authors usually use the
"square-root ansatz":  to take the
square root of the standard model CKM matrix ($V_{CKM}=U_L^+ D_L$)
as an indication of the size of realistic mixings.
It should be denoted that the square root
ansatz must be modified because of the strong constraints from the
data of $B^0 - \overline{B^0}$ mixing \cite{buchalla96b}. By taking into account
the experimental constraints, we naturally set  $D_L^{bd}=V_{td}/2$,  and
$D_L^{bs}=V_{ts}/2$ and $D_R=0$ in the numerical calculations\cite{epjc18}.
Under this assumption, only the charged top-pions $\pcc$ and the charged
technipions $\paa$ and $\pbb$ contribute to the
inclusive charmless decays $b \to s \bar{q} q,\; d \bar{q} q$ with $q\in \{ u,d,s\}$
through the strong and electroweak penguin diagrams, and the  top-pion $\pcc$
dominates the new physics contributions within the reasonable parameter space.

Based on previous studies\cite{epjc18}, the data of $B \to X_s \gamma$ decay result in strong
constraint on TC2 model, specifically on the mass of top-pions:
\beq
140 GeV \lesssim  \mpcc \lesssim 220 GeV, \label{eq:limit1}
\eeq
for $\mu=m_b/2 - 2 m_b$, $\epsilon=0.05\pm 0.03$.

In the numerical calculations, we use the same input parameters of the TC2 model
as being used in Ref.\cite{epjc18}:
\beq
\fpit &=&50 GeV, \; F_{\pi}=120 GeV,\; \epsilon=0.05,  \non
\mpaa&=& 100GeV,\; \mpbb=200GeV, \; \mpcc=200^{+20}_{-50} GeV, \label{eq:limit2}
\eeq
where $F_{\pi}$ and $\fpit$ are the decay constants for technipions and
top-pions, respectively.

\section{ Effective Wilson coefficients } \label{sec:3}

The standard theoretical frame to calculate the inclusive three-body decays
$b \to s \bar{q} q $  \footnote{For $b \to d \bar{q} q$ decays, one simply make the
replacement $s \to d$ properly .} is based on the effective Hamiltonian
\cite{ali9804},
\beq
{\cal H}_{eff}(\Delta B=1) = \frac{G_F}{\sqrt{2}} \left \{
\sum_{j=1}^2 C_j \left ( V_{ub}V_{us}^* Q_j^u  + V_{cb}V_{cs}^* Q_j^c \right )
- V_{tb}V_{ts}^* \left [ \sum_{j=3}^{10}  C_j Q_j  + C_{g} Q_{g} \right ]
\right \}, \label{heff2}
\eeq
where the operator $Q_1$ and $Q_2$ are current-current operators, $Q_3 - Q_6$
are QCD penguin operators induced by gluonic penguin diagrams, and the
operators $Q_7 - Q_{10}$ are generated by electroweak
penguins and box diagrams. The operator $Q_{g}$ is the chromo-magnetic dipole
operator generated from the
magnetic gluon penguin. Following Ref.\cite{ali9804}, we also neglect
the effects of the electromagnetic penguin operator $Q_{7\gamma}$, and do not
consider the effect of the weak annihilation and exchange diagrams.

The new strong and electroweak penguin diagrams can be obtained from the
corresponding penguin diagrams in the SM by replacing the internal $W^{\pm}$ lines with the
unit-charged scalar ($\paa, \pbb$ and $\pcc$ ) lines, as shown in Fig.1 of Ref.\cite{epjc18}.
The new physics will manifest itself by modifying the corresponding Inami-Lim functions $C_0(x),
D_0(x), E_0(x)$ and $E'_0(x)$ which determine the coefficients $C_3(\mw), \ldots, C_{10}(\mw)$
and $C_{g}(\mw)$. These modifications, in turn, will change for example the standard
model predictions for the branching ratios and CP-violating asymmetries
for $B \to VV$ decays. For the sake of simplicity, we here use the results as given
in Ref.\cite{epjc18} directly. For more details of the analytical evaluations of
new Feynman diagrams, one can see Ref.\cite{epjc18}.

By using QCD renormalization group equations\cite{buchalla96a}, it is
straightforward to run Wilson coefficients $C_i(\mw)$ from the scale $\mu = O( \mw)$
down to the lower scale $\mu =O(m_b)$. Working consistently  to the NLO precision,
the Wilson coefficients $C_i$ for $i=1,\ldots,10$ are needed in NLO precision,
while it is sufficient to use the leading logarithmic value for $C_{g}$.
The NLO Wilson coefficients are renormalization
scale and scheme dependent, but such dependence will be cancelled by the
corresponding dependence in the matrix elements of the operators in ${\cal H}_{eff}$.
In the NDR scheme and for $SU(3)_C$, the effective
Wilson coefficients $C_i^{eff}$ can be written as \cite{ali9804,chen99},
\beq
C_i^{eff} &=& \left [ 1 + \frac{\alpha_s}{4\pi} \, \left(
 \gamma^{(0) T} \log \frac{m_b}{\mu}+\hat{r}^T \right) \right ]_{ij}\,C_j\non
&& +\frac{\alpha_s}{24\pi} \, A_i \left (C_t + C_p + C_g \right)
+ \frac{\alpha_{em}}{8\pi}\, B_i C_e ~, \label{eq:wceff}
\eeq
where $A_i=(0,0,-1,3,-1,3,0,0,0,0)^T$, $B_i=(0,0,0,0,0,0,1,0,1,0)^T$,
the matrices  $\hat{r}_V$ and $\gamma_V$ contain the process-independent
contributions from the vertex diagrams. The anomalous dimension matrix
$\gamma_V$ has been given explicitly, for example, in Refs.\cite{chen99,epjc18}.
The explicit expressions of functions $C_t$, $C_p$, and $C_g$ in Eq.(\ref{eq:wceff})
can be found in previous paper \cite{epjc18}. Following Refs.\cite{ali9804,chen99}, we use
$k^2=m_b^2/2 \pm 2$ \footnote{The quantity $k^2$ is the momentum squared transferred by
the gluon, photon or $Z$ to the $q \overline{q}$ pair in inclusive
three-body decays $b \to s q \bar{q}$ and $d q \bar{q}$ with $q=u,d,s$.}
in the numerical calculation.
In fact, branching ratios considered here are not sensitive to
the value of $k^2$ within the range of $k^2=m_b^2/2 \pm 2$, but the CP-violating
asymmetries are sensitive to the variation of $k^2$.

\section{ Branching ratios of $B \to VV$ decays} \label{sec:br}

With the factorization ansatz \cite{bsw87,ali9804,chen99}, the
decay amplitude $<XY|H_{eff}|B>$ can be factorized into a sum of products of two
current matrix elements $<X|J_1^\mu|0>$ and $<Y|J_{2\mu}|B>$ ( or $<Y|J_1^\mu|0>$ and
$<X|J_{2\mu}|B>$). For $B \to VV$ decays, one needs to evaluate the helicity matrix
element $H_\lambda = <V_1(\lambda) V_2(\lambda)|H_{eff}|B)>$ with
$\lambda=0, \pm 1$. In the B-rest frame, the branching ratio of the decay $B \to V_1 V_2$
is given in terms of $H_\lambda$ by
\beq
{\cal  B}(B \to V_1 V_2 ) &=&  \tau_{B}\,
\frac{|p|}{8\pi M_B^2}\left ( |H_0|^2 + |H_{+1}|^2 + |H_{-1}|^2 \right ),
\eeq
where $\tau(B_u^-)=1.674 ps$ and $\tau(B_d^0)=1.542 ps$ \cite{pdg02}, $|p|$ is the magnitude
of momentum of particle $V_1$ and $V_2$ in the B rest frame
\beq
|p| =\frac{1}{2M_B}\sqrt{[M_B^2 -(M_{1} + M_{2})^2] [ M_B^2 -(M_{1}-M_{2})^2 ]}
\label{eq:pxy},
\eeq
where $M_B$  and $M_{i}$ ($i=1,2$) are the masses of B meson and $V_{i}$ vector meson.
The  three independent helicity amplitudes $H_0$, $H_{+1}$ and $H_{-1}$ can be expressed
by three invariant amplitudes $a, b, c$ defined by the decomposition
\beq
H_\lambda &=& i\epsilon^\mu(\lambda)\eta^\nu(\lambda)\left[
    a g_{\mu\nu}+\frac{b}{M_1 M_2}p_\mu
    p_\nu + \frac{ic}{M_1 M_2}\epsilon_{\mu\nu\alpha\beta}p_1^\alpha
    p^\beta \right], \label{eq:hl}
\eeq
where $p_{i}$ ($i=1,2$) is the four momentum of $V_{i}$, and $p=p_1 + p_2 $ is the
four-momentum of B meson, and
\beq
H_{\pm 1} = a \pm c \sqrt{x^2-1}, ~~~~ H_0 = -ax - b\left ( x^2-1 \right ), \label{eq:h01}
\eeq
with
\beq x = \frac{M_B^2-M_1^2-M_2^2}{2M_1 M_2}
\eeq
For individual decay mode, the coefficients $a, b$ and $c$ can be determined
by comparing the helicity amplitude $H_\lambda = <V_1(\lambda) V_2(\lambda)
|H_{eff}|B)>$ with the expression (\ref{eq:hl}).

In the generalized factorization ansatz \cite{ali9804,chen99}, the
effective Wilson coefficients
$C_i^{eff}$ will appear in the decay amplitudes in the combinations,
\beq
a_{2i-1}\equiv C_{2i-1}^{eff} +\frac{{C}_{2i}^{eff}}{N_c^{eff}}, \ \
a_{2i}\equiv C_{2i}^{eff}     +\frac{{C}_{2i-1}^{eff}}{N_c^{eff}}, \ \ \
(i=1,\ldots,5),  \label{eq:ai}
\eeq
where the effective number of colors $N_c^{eff}$ is treated as a
free parameter varying in the range of $2 \leq N_c^{eff} \leq \infty$,
in order to model the nonfactorizable contribution to the hadronic matrix elements.
We here will not  consider the possible effects of final state
interaction (FSI) and the contributions from annihilation channels
although they may play a significant rule for some $B \to VV$ decays.

In numerical calculations, one usually uses two sets of form factors at the zero momentum
transfer from the  Baner, Stech and Wirbel (BSW) model
\cite{bsw87}, as well as the Lattice QCD and Light-cone QCD sum rules (LQQSR),
respectively. Since the differences induced by using two different sets of form factors are
small when compared with that of the new physics contributions, we here use the BSW form factors
only and list them in Appendix. In the following numerical calculations, we use the decay
amplitudes as given in Appendix A of Ref.\cite{ali9804} directly without further
discussions about the details.

In Table \ref{bvv1}, we present the numerical results of the
branching ratios for the nineteen $B \to VV $ decays in the framework of the SM and TC2 model.
The branching ratios are the averages of the branching ratios
of $B$ and anti-$B$ decays.
The theoretical predictions are made by using the central values
of input parameters as given in Appendix, and assuming $\mpcc=200$GeV and
$N_c^{eff}=2, 3, \infty$ in  the GF approach.
The $k^2$-dependence of the branching ratios is weak in the range of $k^2=m_b^2/2\pm 2\; GeV^2$
and hence the numerical results are given by fixing $k^2=m_b^2/2$.

Following Ref.\cite{ali9804}, the nineteen decay channels under study are also
classified into five classes as specified in the second column of Table \ref{bvv1}.
The first three kinds of decays are tree-dominated. The amplitudes of the class-IV
decays involve one (or more) of the dominant penguin coefficients $a_{4,6,9}$ with
constructive interference among them and these decays are generally $ \nceff $ stable.
The class-V decays, however, are generally not stable against $\nceff$
since the amplitudes of these decays involve large and delicate cancellations due to
interference between strong $\nceff$-dependent coefficients $a_3, a_5, a_7$, and $a_{10}$
and the dominant penguin coefficients $a_4,a_6, a_9$.

Among the nineteen decay modes, only $B \to K^{*+} \phi$, $K^{*0}\phi$ and
$B \to \rho^+ \rho^0$ decays are measured experimentally
\cite{prl86-3718,prl87-151801,belle-0113,belle-0255},
\beq
{\cal B}(B^0 \to  K^{*0} \phi ) &=&  ( 9.8 \pm 2.2 ) \times 10^{-6}
\ \  {\rm [weighted-average \cite{prl86-3718,prl87-151801,belle-0113}]}, \label{eq:brvv1} \\
{\cal B}(B^+ \to  K^{*+} \phi ) &=&
( 10.0 \pm 3.7 )\times 10^{-6} \ \  {\rm [weighted-average \cite{prl86-3718,prl87-151801}]},
\label{eq:brvv2}
\eeq
and
\beq
{\cal B}(B \to \rho^+ \rho^0)&=& (38.5\pm 10.9(stat.) ^{+5.9}_{-5.4}(syst.) ^{+2.5}_{-7.5} (pol.))
\times 10^{-6} \ \ {\rm [Belle\cite{belle-0255}] },
\eeq
where the third error is the error associated with the helicity-mix
uncretainty\cite{belle-0255}. The available upper limits ($90\%$ C.L.) on other decay modes
are taken directly from Ref.\cite{pdg02}.

For the measured $B \to K^{*0} \phi$ and $K^{*+} \phi$ decays, the theoretical predictions
in the SM and TC2 model have a strong dependence on the value of $\nceff$, as illustrated in
Figs.\ref{fig:fig1} and \ref{fig:fig2} where the solid and short-dashed curves show the
theoretical predictions in the SM and the TC2 model, respectively. The data clearly
prefer the range of $3 \lesssim \nceff \lesssim 5$.

For the branching ratio ${\cal B}(B \to \rho^+ \rho^0)$, the SM prediction is
$(8.7 - 16.2)\times 10^{-6}$ for $2 \leq \nceff \leq \infty$, as illustrated in
Fig.\ref{fig:fig3}, which is much smaller
than the first measurement as reported by Belle Collaboration\cite{belle-0255}. The new physics
contribution is also negligibly small: less than $1\%$ with respect to the SM prediction. Of
cause, the Belle's measurement still has a large uncertainty and need to be confirmed by further
measurements.

\section{CP asymmetries in $B\to VV$ decays} \label{sec:acp}

In TC2 model, no new weak phase has been introduced through the interactions
involving new particles and hence the mechanism of CP violation in TC2 model
is the same as in the SM. But the CP-violating asymmetries $\acp$
may be changed by the inclusion of
new physics contributions through the interference between the ordinary
tree/penguin amplitudes in the SM and the new strong and electroweak penguin
amplitudes in TC2 model.

For charged B decays the direct CP violation is defined as
\beq
{\cal  A}_{CP} = \frac{\Gamma(B^+ \to f) -\Gamma(B^- \to \bar{f})}{
\Gamma(B^+ \to f) + \Gamma(B^- \to \bar{f})}\label{eq:acpp}
\eeq
in terms of partial decay widths.

For neutral $B^0 (\bar{B}^0)$ decays, the time dependent CP
asymmetry for the decays of states that were tagged as pure $B^{0}$ or
$\bar{B}^0$ at production is defined as
\beq
{\cal  A}_{CP}(t) = \frac{\Gamma(B^0(t) \to f )
 -\Gamma(\bar{B}^0(t) \to \bar{f} )}{
\Gamma(B^0(t) \to f) + \Gamma(\bar{B}^0(t) \to \bar{f})}\label{eq:acp0}
\eeq
According to the characteristics of the final states $f$, neutral $B\to VV$ decays can be
classified into three classes as described in \cite{ali9805}.
For case-1, $f$ or $\bar{f}$ is not a common final state
of $B^0$ and $\bar{B}^0$, and the CP-violating asymmetry is independent of time. We
use Eq.(\ref{eq:acpp}) to calculate the CP-violating asymmetries for CP-class-1
decays: the charged B and case-1 neutral B decays.

For CP-class-2  (class-3) B decays where $\obar{B^0} \to (f = \bar{f})$ with
$f^{CP}=\pm f $ ( $f^{CP}\neq \pm f $ ), the time-dependent and
time-integrated CP asymmetries are of the form
\beq
{\cal  A}_{CP}(t)&=&\aepa \cos(\Delta m\; t) + \aepb \sin(\Delta m\; t),
\label{eq:acptd}\\
{\cal  A}_{CP}&=&\frac{1}{1+x^2}\aepa + \frac{x }{1 + x^2} \aepb ,
\label{eq:acpti}
\eeq
where $\Delta m =m_H -m_L$ is the mass difference between mass eigenstates
$|B^0_H>$ and $|B^0_L>$, $x=\Delta m/\Gamma \approx 0.755$ for the case of
$B_d^0-\bar{B}_d^0$ mixing \cite{pdg02}, and
\beq
\aepa&=& \frac{1- |\lambda_{CP}|^2}{1 + |\lambda_{CP}|^2}, \ \
\aepb= \frac{-2 Im(\lambda_{CP})}{1 + |\lambda_{CP}|^2}, \label{eq:aep}\\
\lambda_{CP}&=&\frac{V^*_{tb}V_{tq}}{V_{tb}V^*_{tq}}
\frac{ <f|H_{eff}|\bar{B}^0> }{ <f|H_{eff}|B^0>},  \label{eq:lambda}
\eeq
for $b \to q$ transitions.

In  Table \ref{acpvv}, we present  numerical results of $\acp(B \to VV) $
in the SM and TC2 model for $\mpcc=200$ GeV and $\nceff=2,3,\infty$, respectively.
We show the numerical results for the
case of using BSW form factors only since the form factor dependence is weak.
In second column of Table \ref{acpvv}, the roman number and arabic
number denotes the classification of the $B \to VV$ decays using
$\nceff$-dependence and the CP-class for each decay mode as defined in
\cite{ali9804,ali9805}, respectively. The first and second errors of the SM
predictions are the dominant errors induced by the uncertainties of $k^2$ and
the CKM angle $\gamma$. For most decay modes, the new physics corrections are
small in size and therefore will be covered by large theoretical uncertainties.

Very recently, BaBar Collaboration reported their first measurements of CP-violating
asymmetries  for the pure penguin decays $B^0 \to K^{*0} \phi$ and $B^\pm \to K^{*\pm}\phi$
\cite{prd65-051101},
\beq
\acp{(\obar{B^0} \to \obar{K^{*0}} \phi)} = 0.00 \pm 0.27 \pm 0.03, \\
\acp{(B^\pm \to K^{*\pm} \phi)} = -0.43^{+0.36}_{-0.30} \pm 0.06
\eeq
These results  are clearly  consistent with the SM prediction: $\acp(B\to K^* \phi) \sim -2\%$.
The new physics corrections from the new penguin diagrams appeared in the TC2 model can
change the sign of $\acp(B^+ \to K^{*+}\phi)$, but its size is still around $5\%$ for
$\nceff \sim (3-5)$ as indicated by the measured branching ratios.
It is hard to measure such small difference experimentally in near future.
For the CP asymmetry of $B \to K^{*0}\phi$ decay, the new physics correction is
only about $10\%$ with respect to the SM prediction and can be neglected.

For the decays $B \to \rho^0 \rho^0$ and $\omega \omega$, although their CP asymmetries
can be large in size in both the SM and TC2 model, but these decays are not promising
experimentally because of their very small branching ratios ($\sim 10^{-7}$).

\section{Summary} \label{sec:sum}

In this paper, we calculated the branching ratios
and CP-violating asymmetries of the nineteen $B \to VV$ decays in the SM and the TC2 model
by employing the effective Hamiltonian with GF approach.

\begin{figure}[]
\vspace{-40pt}
\begin{minipage}[]{\textwidth}
\centerline{\epsfxsize=0.9\textwidth \epsffile{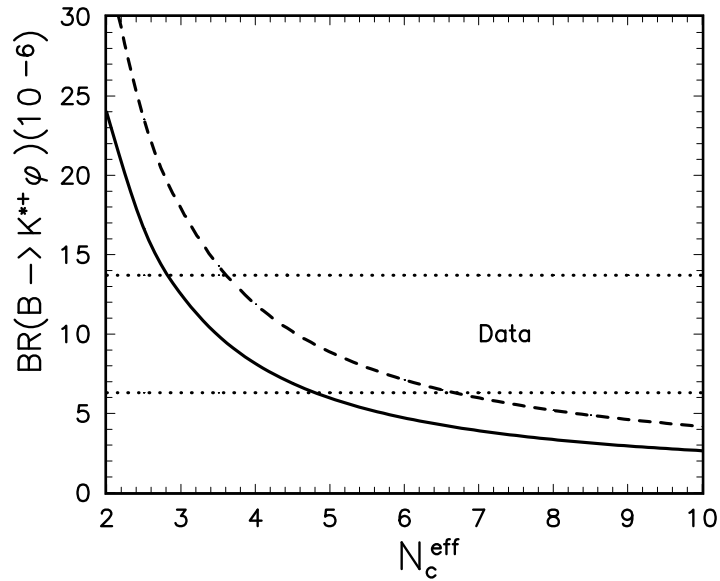}}
\vspace{-20pt}
\caption{Plots of the branching ratio ${\cal B}(B \to K^{*+} \phi)$ versus
$\nceff$ in the SM (solid curve) and the TC2 model (short-dashed curve) for
$\mpcc=200$ GeV and $\nceff=2-10$. The band between two dots lines corresponds
to the data ${\cal B}(B \to K^{*+} \phi)=(10.0 \pm 3.7)\times 10^{-6}$.}
\label{fig:fig1}
\end{minipage}
\end{figure}

\begin{figure}[]
\vspace{-40pt}
\begin{minipage}[]{\textwidth}
\centerline{\epsfxsize=0.9\textwidth \epsffile{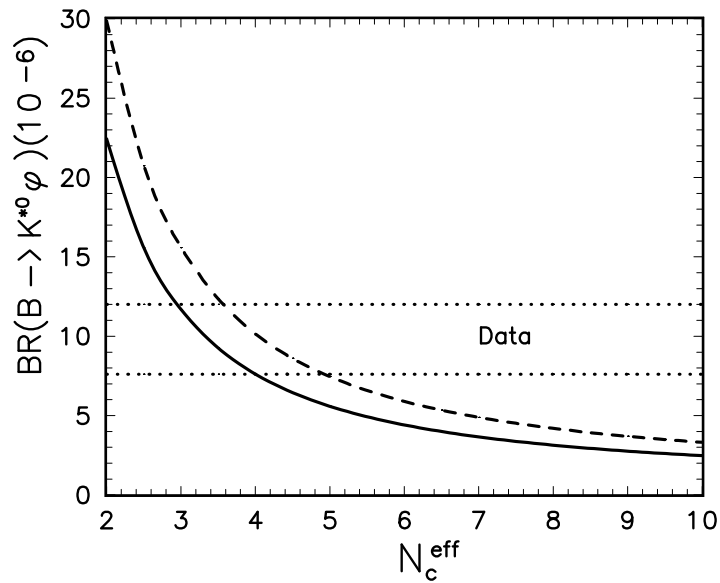}}
\vspace{-20pt}
\caption{The same as Fig.1 but for ${\cal B}(B \to K^{*0} \phi)$ decay.
The band between two dots lines corresponds
to the data ${\cal B}(B \to K^{*0} \phi)=(9.8 \pm 2.2)\times 10^{-6}$.}
\label{fig:fig2}
\end{minipage}
\end{figure}

\begin{figure}[]
\vspace{-40pt}
\begin{minipage}[]{\textwidth}
\centerline{\epsfxsize=0.9\textwidth \epsffile{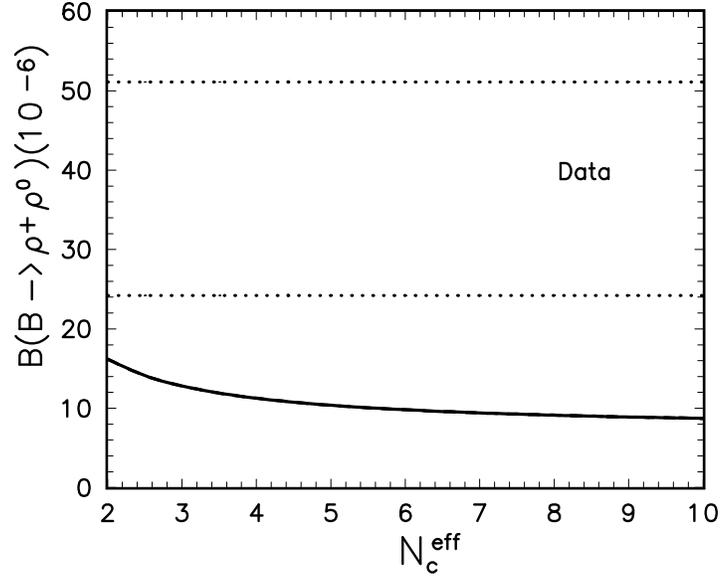}}
\vspace{-20pt}
\caption{Plots of the branching ratio ${\cal B}(B \to \rho^+ \rho^0)$ versus
$\nceff$ in the SM and the TC2 model for $\mpcc=200$ GeV and $\nceff=2-10$, but two curves
coincide each other. The band between two dots lines corresponds
to the Belle's measurement: ${\cal B}(B \to \rho^+ \rho^0)=(38.5^{+12.6}_{-14.3})
\times 10^{-6}$.}
\label{fig:fig3}
\end{minipage}
\end{figure}

In section \ref{sec:br} we presented the numerical results for the branching
ratios ${\cal B}(B \to VV)$ in Table \ref{bvv1} and displayed the  $\nceff$-dependence
of the three measured decay modes in Figs.\ref{fig:fig1},\ref{fig:fig2} and \ref{fig:fig3}.
From these table and figures, the following conclusions can be reached.

The theoretical predictions in the SM and the TC2 model for all nineteen decay modes
are well consistent with currently available experimental measurements and upper
limits within errors.
For tree-dominated decays, the new physics enhancements are usually small. For
penguin-dominated decays, such as $B \to K^{*+}\phi$ and $K^{*0}\phi$ decays,
the new physics enhancements to the branching ratios are around $40\%$.
The theoretical uncertainties induced by varying $k^2$, $\gamma$ and $\mpcc$
are moderate within the range of $k^2 = m_b^2/2 \pm 2 GeV^2$, $\gamma=(70 ^{=10}_{-20})^\circ$
and $\mpcc=200^{+20}_{-50}$ GeV. The $\nceff$-dependence vary greatly for different decay modes.

From Figs.\ref{fig:fig1} and \ref{fig:fig2}, one can see that the measured branching ratios
${\cal B}(B^+ \to K^{*+} \phi)$ and ${\cal B}(B^0 \to K^{*0} \phi)$ prefer the
range of $3 \lesssim \nceff \lesssim 5$. For the branching ratio ${\cal B}(B^+ \to \rho^+
\rho^0)$, however, the SM and TC2 model predictions are the same and less than
half of the first measurement reported by Belle Collaboration\cite{belle-0255} as illustrated in
Fig.\ref{fig:fig3}.

\begin{figure}[]
\vspace{-40pt}
\begin{minipage}[]{\textwidth}
\centerline{\epsfxsize=0.9\textwidth \epsffile{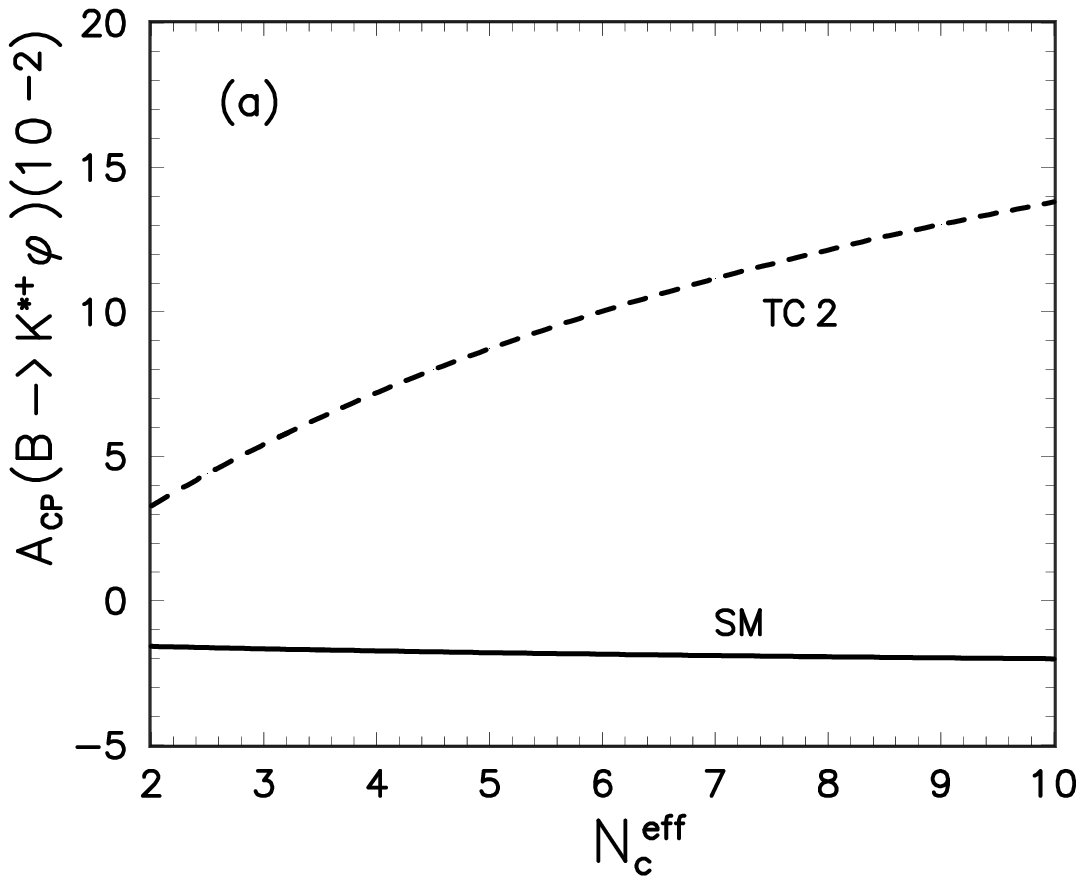}}
\vspace{-40pt}
\centerline{\epsfxsize=0.9\textwidth \epsffile{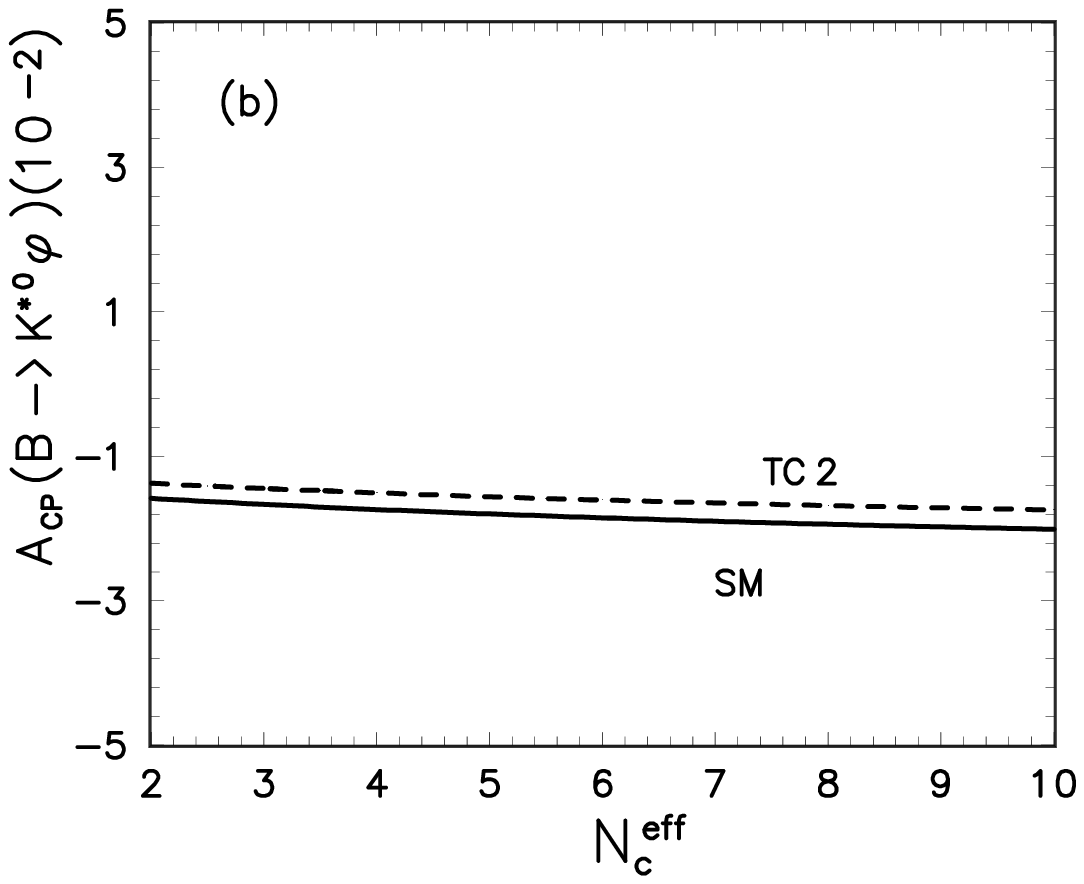}}
\vspace{-20pt}
\caption{Plots of the CP asymmetries of $B \to K^{*+}\phi$ and
$B \to K^{*0}\phi$ decays
versus $\nceff$ in the SM (solid curve) and the TC2 model (short-dashed curve)
for $\mpcc=200$ GeV and $\nceff=2-10$. The Babar's limits at $90\%$ C.L. are
$\acp{(B^+ \to K^{*+} \phi)} = [-0.88, +0.18]$  and
$\acp{(B^0 \to K^{*0} \phi)} = [-0.44, +0.44]$.}
\label{fig:fig4}
\end{minipage}
\end{figure}

In section \ref{sec:acp}, we calculated the CP-violating asymmetries $\acp$
for  $B \to VV$ decays in the SM and TC2 model, presented the
numerical results in Table \ref{acpvv} and displayed the
$\nceff$-dependence of $\acp$ for decays $B^+ \to K^{*+} \phi$ and
$B^0 \to K^{*0} \phi $ in Fig.\ref{fig:fig4}.
For most $B \to VV$ decays,  the new physics corrections  are generally
small or moderate in magnitude (  $ < 30\% $), and
insensitive to the variation of $\mpcc$ and $\nceff$. For $B \to K^{*+}\phi$ decay, the
sign of $\acp$ can be changed by including the new physics contributions, but the
theoretical predictions in both the SM and TC2 model are still well consistent with the data.

\section*{ACKNOWLEDGMENTS}
Authors acknowledge the support by the National Natural Science Foundation of
China under Grant No.~10075013 and 10275035, and by the Research Foundation of
Nanjing Normal University under Grant No.~214080A916.

\vspace{1cm}

\section*{Appendix: Input parameters and form factors} \label{app:a}

In this Appendix we list all input parameters and form factors used in
this paper.

\begin{itemize}

\item
Coupling constants and masses( in unit of GeV ), $\cdots$,
\beq
\alpha_{em}&=&1/128, \;  \alpha_s(M_Z)=0.118,\;  \sin^2\theta_W=0.23,\;
G_F=1.16639\times 10^{-5} (GeV)^{-2}, \non
M_Z&=&91.188, \;   M_W=80.42,\;
m_{B_d^0}=m_{B_u^\pm}=5.279,\;   m_{\rho}=0.770,\;  m_{\omega}=0.782,\non
m_{\phi}&=&1.019,\; m_{K^{*\pm}}=0.892,\; m_{K^{*0}}=0.896,\label{masses}
\eeq

\item
Wolfenstein paramters:
\beq
A=0.847,\; \lambda=0.2205, \; R_b=0.38, \gamma={(70 ^{+10}_{-20})}^\circ.
\eeq

\item
The current quark masses $m_i$ ($i=u,d,s,c,b$, and $\mu=2.5 GeV$)
\beq
m_b=4.88 GeV, \; m_c=1.5 GeV, m_s=0.122 GeV, \; m_d=7.6 MeV,\; m_u=4.2 MeV.
\label{cur-mass}
\eeq
For the mass of heavy top quark we also use $m_t=\overline{m_t}(m_t)=168 GeV$.

\item
The decay constants of light mesons:
\beq
f_{K^*}=214 MeV, \; f_{\rho}=210 MeV, \; f_{\omega}=195 MeV, \; f_{\phi}=233 MeV. \label{fpis}
\eeq

\item
The form factors at the zero momentum transfer in the Baner, Stech and Wirbel (BSW)
\cite{bsw87} model as given in Ref.\cite{ali9804}:
\beq
&&A_{0,1,2}^{B\rho}(0)=A_{0,1,2}^{B\omega}(0)=0.28,\;
A_{0}^{BK^*}(0)=0.32, \;A_{1,2}^{BK^*}(0)=0.33, \non
&&V^{B\rho}(0)=V^{B\omega}(0)=0.33, \; V^{BK^*}(0)=0.37. \label{eq:bsw-f}
\eeq
The $k^2$-dependence of the form factors were defined in Ref.\cite{bsw87} as
\beq
A_0(k^2)&=& \frac{A_0(0)}{1-k^2/m^2(0^-)}, \ \
A_1(k^2)=   \frac{A_1(0)}{1-k^2/m^2(1^+)},  \non
A_2(k^2)&=& \frac{A_2(0)}{1-k^2/m^2(1^+)}, \ \
V(k^2) =    \frac{V(0)}{1-k^2/m^2(1^-)}.
\eeq

\item
The pole masses (in unit of GeV) being used to evaluate the $k^2$-dependence of form factors
are,
\beq
\{ m(0^-),m(1^-),m(1^+),m(0^+) \}&=& \{ 5.2789, 5.3248,5.37,5.73 \}
\eeq
for $\bar{u}b$ and $ \bar{d}b$ currents. And
\beq
\{ m(0^-),m(1^-),m(1^+),m(0^+)\}&=& \{5.3693, 5.41,5.82,5.89\}
\eeq
for $\bar{s}b $ currents.

\end{itemize}

\newpage
\begin{table}[]
\begin{center}
\caption{$B\to V V$ branching ratios (in units of $10^{-6}$) using the BSW
form factors in the SM and TC2,  with $k^2=m_b^2/2$, $A=0.847$, $\lambda=0.2205$, $R_b=0.38$,
$\gamma=70^\circ$, $\mpcc=200$ GeV and  $\nceff= 2,\; 3,\; \infty$.
The last column contains the experimental measurements for $B \to K^{*0} \phi$ and
$K^{*+} \phi$ decays, and the upper limits ($90\%$ C.L.) on other decay modes taken from
PDG tables [27].}
\label{bvv1}
\vspace{0.2cm}
\begin{tabular} {l|l|c|c|c|c|c|c|l} \hline
 & & \multicolumn{3}{c|}{ SM }& \multicolumn{3}{c|}{ TC2 model }  & Data \\ \cline{1-9}
Channel                        & Class &  $2$ & $3$ & $\infty$ & $2$ & $3$ & $\infty$  &   \\ \hline
$ B^0 \to \rho^+ \rho^-$       &I  &$ 21.2 $&$  24.1$ &$30.5$& $ 21.4$&$ 24.3$ &$ 30.6$& $<2200$\\
$ B^0 \to \rho^0 \rho^0 $      &II &$ 0.47 $&$  0.10$ &$1.85$& $ 0.49$&$ 0.12$ &$ 1.88$& $<18$ \\
$ B^0 \to \omega \omega$       &II &$ 0.91 $&$  0.17$ &$1.46$& $ 1.08$&$ 0.22$ &$ 1.46$& $<19$ \\
$ B^+ \to \rho^+ \rho^0 $      &III&$ 16.2 $&$  12.8$ &$7.20$& $ 16.2$&$ 12.8$ &$ 7.20$& $38.5^{+12.4}_{-14.3} $ \\
$ B^+ \to \rho^+ \omega$       &III&$ 17.3 $&$  13.8$ &$7.90$& $ 18.0$&$ 14.1$ &$ 7.92$& $<61$ \\
$ B^0 \to K^{*+} \rho^- $      &IV &$ 6.65 $&$  7.50$ &$9.32$& $ 10.5$&$ 11.4$ &$ 13.3$& $-$ \\
$ B^0 \to K^{*0} \rho^0 $      &IV &$ 1.96 $&$  2.29$ &$3.01$& $ 2.29$&$ 2.69$ &$ 3.73$&  $<34$\\
$ B^+ \to K^{*+} \rho^0 $      &IV &$ 5.89 $&$  6.59$ &$8.57$& $ 10.1$&$ 11.5$ &$ 14.8$&  $<74$\\
$ B^+ \to K^{*0} \rho^+ $      &IV &$ 7.41 $&$  9.31$ &$13.8$& $ 11.3$&$ 14.4$ &$ 21.8$&  $-$\\
$ B^+ \to K^{*+} \bar{K}^{*0}$ &IV &$ 0.42 $&$  0.53$ &$0.78$& $ 0.64$&$ 0.81$ &$ 1.22$&  $<71$\\
$ B^0 \to K^{*0} \bar{K}^{*0}$ &IV &$ 0.39 $&$  0.49$ &$0.73$& $ 0.60$&$ 0.76$ &$ 1.15$&  $<22$\\
$ B^0 \to \rho^0 \omega$       &V  &$ 0.48 $&$  0.26$ &$0.02$& $ 0.64$&$ 0.34$ &$ 0.03$&  $<11$\\
$ B^0 \to K^{*0} \omega$       &V  &$ 14.6 $&$  4.90$ &$1.12$& $ 19.9$&$ 6.73$ &$ 1.39$& $<23$ \\
$ B^+ \to K^{*+} \omega$       &V  &$ 14.8 $&$  4.32$ &$3.14$& $ 20.8$&$ 5.94$ &$ 4.36$&  $<87$\\
$ B^+ \to K^{*+} \phi$         &V  &$ 24.3 $&$  12.6$ &$0.67$& $ 33.9$&$ 18.1$ &$ 1.26$&$10.0 \pm 3.7$ \\
$ B^0 \to K^{*0} \phi$         &V  &$ 22.3 $&$  11.6$ &$0.61$& $ 29.8$&$ 15.5$ &$ 0.83$&$9.8 \pm 2.2$  \\
$ B^+ \to \rho^+ \phi$         &V  &$ 0.07 $&$  0.004$&$0.52$& $ 0.06$&$ 0.02$ &$ 0.85$& $<16$ \\
$ B^0 \to \rho^0 \phi$         &V  &$ 0.03 $&$  0.002$&$0.24$& $ 0.03$&$ 0.01$ &$ 0.39$& $<13$ \\
$ B^0 \to \omega \phi$         &V  &$ 0.03 $&$  0.002$&$0.24$& $ 0.03$&$ 0.01$ &$ 0.39$& $<21$ \\
\hline
\end{tabular}\end{center}
\end{table}

\begin{table}[]
\begin{center}
\caption{ CP-violating asymmetries $\acp$ of $B \to VV $ decays
(in percent) in the SM  and TC2 model
for   $k^2=(m_b^2/2 \pm 2)GeV^2$, $\gamma={(70^{+10}_{-20})}^\circ$,
$\mpcc=200$ GeV and $\nceff=2,3,\infty$. }
\label{acpvv}
\vspace{0.2cm}
\begin{tabular} {l|l|c|c|c|c|c|c} \hline
 & &\multicolumn{3}{c|}{ SM }& \multicolumn{3}{c }{ TC2 model }  \\ \cline{1-8}
Channel                              &Class&$2$&$3$&$\infty$&$2$&$3$&$\infty$\\ \hline
$ \obar{B^0} \to \rho^\pm \rho^\mp$  &I-3  &$ 9.8^{+0.1 +27.7}_{-0.7 -14.8}$&$  9.7^{+0.2 +27.8}_{-0.7 -14.7}$ &$  9.6^{+0.1+ 27.8}_{-0.7 -14.8} $& $   12.8$&$  12.5$ &$   11.9 $ \\
$ \obar{B^0} \to \rho^0 \rho^0 $     &II-3 &$-45.5^{+5.5 +30.6}_{-1.7 -6.6}$&$  18.7^{+5.6 +1.1}_{-4.3 -1.3} $ &$  28.9^{+0.5 +18.2}_{-1.4 -11.4}$& $   12.8$&$  12.5$ &$   11.9 $  \\
$ \obar{B^0} \to \omega \omega$      &II-3 &$ 60.0^{+1.6 +1.8}_{-3.1 -5.2} $&$  18.0^{+5.3 +0.8}_{-2.9 -1.2} $ &$  13.8^{+0.1 +25.6}_{-0.7 -14.3}$& $   58.4$&$  15.3$ &$   15.5$  \\
$ B^\pm \to \rho^\pm \rho^0 $        &III-1&$ 0.2  \pm 0.1 \pm 0.0         $&$  0.2\pm 0.1 \pm 0.0           $ &$  0.3^{+0.0 +0.1}_{-0.1 -0.0}   $& $   0.2 $&$  0.2 $ &$   0.3$  \\
$ B^\pm \to \rho^\pm \omega$         &III-1&$ 8.6^{+2.0 +1.3}_{-4.2 -2.5}  $&$  7.5^{+1.7 +0.9}_{-3.8 -2.1}  $ &$  3.9^{+1.0 +0.3}_{-2.2-0.9}    $& $   9.0 $&$  7.9 $ &$   4.2$  \\
$ \obar{B^0} \to K^{*\pm} \rho^\mp $ &IV-1 &$-14.5^{+8.5 +1.3}_{-4.7 -1.5} $&$ -14.6^{+7.4 +1.3}_{-4.7 -1.5} $ &$ -14.7^{+8.6 +1.3}_{-4.8 -1.5}  $& $  -10  $&$ -10.4$ &$  -11.2$  \\
$ \obar{B^0} \to K^{*0} \rho^0 $     &IV-1 &$ 2.6^{+1.8 +0.3}_{-3.0 -0.6}  $&$ -2.0 \pm 0.1^{+0.3}_{-0.2}    $ &$ -8.9^{+4.7 +0.7}_{-2.8 +0.2}   $& $   2.3 $&$ -1.8 $ &$  -7.8$  \\
$ B^\pm \to K^{*\pm} \rho^0 $        &IV-1 &$-11.7^{+6.7+1.0}_{-3.6 -1.1}  $&$ -9.8^{+5.4 +0.6}_{-3.0 -0.3}  $ &$ -6.5^{+3.4 +0.4}_{-1.8 +0.2}   $& $  -7.3 $&$ -6.2 $ &$  -4.2$  \\
$ B^\pm \to \obar{K^{*0}} \rho^\pm $ &IV-1 &$-1.6 \pm 0.1 ^{+0.3}_{-0.1}   $&$ -1.5 \pm 0.1 ^{+0.3}_{-0.1}   $ &$ -1.4 \pm 0.1 ^{+0.2}_{-0.1}    $& $  -1.3 $&$ -1.2 $ &$  -1.1$  \\
$ B^\pm \to K^{*\pm} \obar{K^{*0}}$  &IV-1 &$ 14.2^{+6.0 +0.2}_{-3.2 -0.9} $&$  13.4^{+5.8 +0.3}_{-3.1 -0.8} $ &$  12.2^{+7.7 +0.4}_{-2.9 -0.7}  $& $   11.2$&$  10.5$ &$   9.5$  \\
$ \obar{B^0} \to K^{*0} \bar{K}^{*0}$&IV-3 &$ 17.6^{+6.1 +0.9}_{-3.3 -1.2} $&$  16.7^{+5.8 +0.8}_{-3.3 -1.2} $ &$  15.4^{+5.6 +0.8}_{-3.0 -1.1}  $& $   14.2$&$  13.4$ &$   12.3$  \\
$ \obar{B^0} \to \rho^0 \omega$      &V-3  &$ 12.0^{+7.0 +0.9}_{-4.0 -1.2} $&$  18.3^{+6.3 +0.8}_{-3.5 -1.3} $ &$  64.4^{+1.9 +1.9}_{-3.2 -7.2}  $& $   10.4$&$  15.9$ &$   59.7$  \\
$ \obar{B^0} \to \obar{K^{*0}}\omega$&V-1  &$ 11.2^{+6.4 +0.9}_{-3.4 -1.0} $&$  12.6^{+5.7 +0.3}_{-3.0 -0.8} $ &$  3.2 ^{+9.5 +3.0}_{-4.9 -1.3}  $& $   9.5 $&$  10.5$ &$   3.1$  \\
$ B^\pm \to K^{*\pm} \omega$         &V-1  &$ 5.6^{+10.5 +3.4}_{-5.6 -1.7} $&$  0.9^{+14.4 +4.8}_{-7.9 -1.7} $ &$  10.6 \pm 0.1^{+1.0}_{-2.1}    $& $   5.1 $&$  1.7 $ &$   8.6$  \\
$ B^\pm \to K^{*\pm} \phi$           &V-1  &$-1.6 \pm 0.1 ^{+0.3}_{-0.1}   $&$ -1.7 \pm 0.1 ^{+0.3}_{-0.1}   $ &$ -2.6  \pm 0.1^{+0.5}_{-0.2}    $& $   3.3 $&$  5.4 $ &$   27.0$ \\
$ \obar{B^0} \to \obar{K^{*0}} \phi$ &V-1  &$-1.6 \pm 0.1 ^{+0.3}_{-0.1}   $&$ -1.7 \pm 0.1 ^{+0.3}_{-0.1}   $ &$ -2.6  \pm 0.1^{+0.5}_{-0.2}    $& $  -1.4 $&$ -1.5 $ &$  -2.3$  \\
$ B^\pm \to \rho^\pm \phi$           &V-1  &$ 13.4^{+5.4 +0.3}_{-3.1 -0.8} $&$  1.0 ^{+0.8}_{-0.4}\pm 0.1    $ &$  9.6 ^{+4.9 +0.4}_{-2.5 -0.6}  $& $   14.4$&$  0.5 $ &$   7.3$  \\
$ \obar{B^0} \to \rho^0 \phi$        &V-3  &$ 16.9^{+5.9 +0.8}_{-3.3 -1.2} $&$  1.5 ^{+0.7}_{-0.4} \pm 0.1   $ &$  12.5^{+4.8 +0.8}_{-2.6 -0.9}  $& $   17.7$&$  0.7 $ &$   9.8$  \\
$ \obar{B^0} \to \omega \phi$        &V-3  &$ 16.9^{+5.9 +0.8}_{-3.3 -1.2} $&$  1.5^{+0.7}_{-0.4} \pm 0.1    $ &$  12.5^{+4.8 +0.8}_{-2.6 -0.9}  $& $   17.7$&$  0.7 $ &$   9.8$  \\
\hline
\end{tabular}\end{center}
\end{table}

\newpage

\newpage

\newpage

\end{document}